**Effect of the structure of lead iodine perovskites on the photovoltaic efficiencies**

César Tablero Crespo
*Instituto de Energía Solar, E.T.S.I. de Telecomunicación,
Universidad Politécnica de Madrid,
Ciudad Universitaria s/n, 28040 Madrid, SPAIN.*
e-mail: ctablero@etsit.upm.es

**Abstract**

Methyl-ammonium lead iodide perovskite crystallizes in different structures depending on the temperature: orthorhombic, tetragonal and cubic. An important point to be considered is the effect of the microscopic properties of the different structures on the optical and photovoltaic properties. Using first principles we obtain the absorption coefficients that will determine the absorption of solar radiation. In order to analyze the contributions of the different atoms to the absorption coefficients we split them into a many-species expansion. Using a similar methodology we also split the efficiencies as a many-species expansion. It allows the contribution of the atomic species to be identified and quantified to the absorption coefficients and to the solar cell efficiency of the different phases. Additionally the effect of the cell thickness $w$ is quantified.

**1. Introduction**

The iodine lead halide perovskites absorb in the spectral region between 1.55-1.61 eV [1,2]. These values are close to the optimum one for photovoltaic performance (~1.3 eV for a sunlight spectrum with the maximum concentration) [3–5]. This bandgap range makes these materials very good candidates to absorb solar radiation in solar cells with efficiencies of 15-20 % [6,7]. Furthermore, these materials present high external radiative efficiency with relatively small difference between the open-circuit voltage (transformed into energy) and the energy bandgap. It indicates low non-radiative recombination rates compared to other thin-film semiconductors. The more efficient solar cell has been obtained from $(MA)PbI_3$ (MAPI) which has a bandgap of 1.55 eV [7], where MA is the methyl ammonium cation ($MA=[CH_3NH_3]^+$).

Organic/inorganic perovskites are a general class of materials normally with the $AMX_3$ chemical formula. $A^+$ is a large organic/inorganic cation, $M^{+2}$ a smaller metal cation (M=Pb, Ge, Sn) and $X^-$ an anion from the halide series (X=I, Br, Cl). Structurally, the A cation resides within the network of corner sharing $MX_6$ octahedral structures having different tilting characteristics at various structural phases of the compound. In this work, the cation A is the MA organic cation, M=Pb, and X=I. Both the optical absorption and the photoluminescence is related mainly to the Pb and I combination.

The MAPI has several structural phases [8]: a cubic (C) phase (space group $Pm\bar{3}m$) above 330 K, a tetragonal (T) phase (space group *I4/mcm*) from 160 to 330 K, and an orthorhombic (O) phase (space group *Pnma*) below 160 K. The MA cations are fully and partially disordered in the C and T phase respectively, whereas in the O phase the cations take a fixed position in the lattice [8,9]. A large number of theoretical and experimental studies [10–15] have focused on the structural evolution across these phases. In some cases other intermediate phases appear between the previous ones. For example, MAPI transforms from tetragonal *I4/mcm* to tetragonal *P4mm* and not directly to the cubic $Pm\bar{3}m$ [10]. MAPI single crystals can be grown by several methods such as antisolvent vapor-assisted crystallization [16], top-seeded solution growth [17], and solution-based hot-casting technique [18].

A number of studies have investigated the impact of the phase transitions on optical [11] and PV photovoltaic properties [9]. Previous studies have shown the absence of large changes in the optical properties [11] and in the photovoltaic parameters across the T→C phase transition. However a strong efficiency decrease when approaching the orthorhombic phase [11] has also been observed.

An important aspect to consider is the relationship between the structural and optical properties and how they affect to the photovoltaic properties. For this, in this work, using *ab initio* methods and different structural phases, we will relate the microscopic contributions, both to the absorption coefficients and to the efficiency.

To analyze the photovoltaic characteristics we will use the absorption coefficients obtained theoretically as a photon energy function. This additionally allows us to quantify the main photovoltaic parameters as a function of the thickness of the absorbent material. This procedure differs from the usual one where step functions are used for the absorption above the semiconductor bandgap and infinite optical thickness is assumed to absorb all photons of solar radiation. In order to highlight the main microscopic contributions and quantify how the structural phase transitions affect the photovoltaic performance we will split the absorption coefficients and efficiencies as a exact many-specie expansion [19,20]. Finally, we will analyze the differences between the structural phases and their effect on the main photovoltaic parameters (currents, voltages and efficiencies). This information is relevant since a significant change in the optical and photovoltaic properties associated with a structural phase transition in the solar cell operating range would drastically modify the main parameters of the solar cell. The results give explanation for the absence of drastic changes in the optical and photovoltaic properties, in particular for T→C phase transition within the solar cell operating regime.

It is clear that the electronic properties are mainly associated with $PbI_6$ octahedral structures. However, we will show that in addition to this effect on the electronic properties, the influence of the organic cation can be very important in the optical properties and on the photovoltaic response.

## 2. Methodology

Calculations were made using density functional theory (DFT) [21,22] with periodic boundary conditions and with the Perdew-Burke-Ernzerhof [23] generalized gradient approximation for the exchange-correlation functional. The interactions of the valence electrons is modeled by the Troullier–Martins [24] pseudopotentials expressed in the Kleinman–Bylander [25,26] form. The valence electrons are represented by a basis set made up of localized pseudoatomic orbitals [27].

The lattice parameters in the literature for the different phases are: $a$=6.391 Å [12]-6.328 Å [8] for the C structure, $(a|c)$=(8.800|12.685) Å [12], (8.86|12.66) Å [8], (8.849|12.642) Å [13] for the T structure, and $(a|b|c)$=(8.836|12.580|8.555) Å [14], (8.861|12.620|8.581) Å [12]. In this work we have used the lattice parameters $a$=6.391 Å, $(a|c)$=(8.859|13.055) Å, and $(a|b|c)$=( 8.844|12.592|8.563) Å for the C, T and O structures. The irreducible Brillouin zone for the electronic properties was sampled using 256 (72) special k points for the C (T and O) structure.

The absorption coefficient $\alpha$ as a photon energy $E$ function is the sum of all possible transitions among the $\mu$ and $\lambda$ band states [28,29] at the **k** points in the Brillouin zone:

$$\alpha(E) \sim \frac{1}{E} \sum_{\lambda} \sum_{\mu > \lambda} \int d\mathbf{k} (f_{\lambda\mathbf{k}} - f_{\mu\mathbf{k}}) |p_{\mu\lambda}|^2 \delta(E_{\mu\mathbf{k}} - E_{\lambda\mathbf{k}} - E) \quad (1)$$

$\alpha(E)$ depend on the band state energies ($E_{\mu\mathbf{k}}$), the ocupation factors ($f_{\mu\mathbf{k}}$), and the transition probabily from $\lambda$ to $\mu$ band states [28,29] proportional to the momentum matrix elements $p_{\mu\lambda}$. All these magnitudes, and the band wave functions, depend on the interactions between the atoms that make up the crystal. Therefore, by varying the relative position of the atoms and the symmetry [30], the interactions and Brillouin zones are different. For this reason, the electronic and optical properties, such as the bandgaps and absorption coefficients, are different in the three perovskite structures even though the atoms are the same.

The momentum matrix elements can be split [19,20] as $p_{\mu\lambda} = \sum_A \sum_B p_{\mu\lambda}^{AB}$ where $p_{\mu\lambda}^{AB}$ is the inter-species (intra-species if A=B) component coupling the basis set functions on different species atoms A and B. For direct transitions the transition probabily is proportional to $|p_{\mu\lambda}|^2$. Note that $|p_{\mu\lambda}|^2$ involve terms like $|p_{\mu\lambda}^{AB}| \cdot |p_{\mu\lambda}^{CD}|$. Therefore the absorption coefficient can be split as an exact many-species expansion [19,20] with terms no larger than those of 4 species: $\alpha = A_{12} + A_{34}$. The first term involves intra-species (A=B) and inter-species (A≠B) contributions $A_{12} = \sum_A \sum_B \alpha_{AB}$. $A_{34}$ involves 3-species (A≠B≠C) and 4-species (A≠B≠C≠D) contributions. Terms of two, three and four species will only be present if the compound has more than one, two, and three species respectively. In binary compounds there is no term of three or four species ($A_{34}$=0 and $\alpha \approx A_{12}$). In compounds with 3 | 4 | 5 different species there are 1 | 4 | 10 terms of three species and 0 | 1 | 5 terms of four species. In the case of this perovskite with 5 different species (C, H, N, Pb, I), the sum of terms of 3 and 4 species $A_{34}$ is 20, greater than the number of terms of 2 species $A_{12}$ (10). The number of $A_{34}$ terms increases rapidly with the number of species of the compound, and therefore its contribution to the absorption coefficient.

Under illumination, with an incident spectrum photon flux density $g_a$, the photons with energy E are absorbed creating carriers (electrons and holes). Assuming that each absorbed photon creates an electron-hole pair (quantum efficiency=1), the current generated by the absorption process [3–5] is $J_a = q \int a(E,\alpha,w) g_a(E) dE$, where $a(E,\alpha,w) \simeq 1 - e^{-\alpha w}$ is the absorptivity, $\alpha(E)$ the absorption coefficient, $w$ the optical thickness, and $q$ is the electron charge. In this work we have used the AM1.5G [31] spectrum for $g_a$. As a result of the radiative recombination, the absorption process is always accompanied by an emission process with current $J_e = q \int a(E,\alpha,w) g_e(E) dE$, where $g_e$ is the photon flux density emitted. It is usual to consider that the solar cell emits radiation as a generalized Black-body at a temperature T=300 K [3–5]. Then $g_e(E) \sim E^2 b(E,T,\mu)$, where $b(E,T,\mu) = [e^{(E-\mu)/kT} - 1]^{-1}$ is the Bose factor, $k$ the Boltzmann constant, $\mu = qV$ the chemical potential associated with the radiation emitted [3–5], and $V$ the voltage. Note that the total current depends on the voltage via the emission current, i.e. $J(V) = J_a - J_e(V)$. Finally the efficiency $\eta$ of a solar cell is $\eta = JV / P_{inc}$ where $P_{inc}$ is the irradiance of the incident sun-light spectrum.

If the absorption coefficient is split as $\alpha = \sum_B \alpha_B$ the current generated by $\alpha_B$ is $J_B = q \int a \cdot (\alpha_B / \alpha) \cdot (g_a - g_e) dE$, and the associated efficiency is $\eta_B = J_B V / P_{inc}$. Obviously $J = \sum_B J_B$ and $\eta = \sum_B \eta_B$. Then if the absorption coefficient is expressed as a many-species expansion [19,20], the currents and efficiencies can also be split into as a many-species expansion. As with the absorption coefficients, this decomposition is exact and there are no terms larger than those of 4 species.

## 3. Results and discussion

### 3.1 Electronic structure

It is known that MAPI undergoes structural phase transitions from O (space group *Pnma*) to T(space group *I4/mcm*) at ~170 K and then to cubic (space group $Pm\bar{3}m$) at ~330 K [8,10,15]. The experimental and theoretical bandgaps for the different structures in the literature are shown in Table 1 together with those obtained in this work. These calculated bandgaps are in general in good agreement with reported experimental values. These results also agree with the experimental red-shift of the bandgap (bandgap decreasing) associated with the O→T and C→T change in the crystal structure [32].

**Table 1**: Bandgaps (eV) for the different C, T, and O structures obtained experimentally, theoretically and those obtained in this work.

|  | C structure | T structure | O structure |
|---|---|---|---|
| Experimental | 1.69 [32] | 1.61 [32] -1.63 [33,34] | 1.65 [32] -1.66 [35] |
| Theoretical | 1.16-1.28 [32], 1.30 eV [14] | 1.43 [14], 1.67 [32], 0.98-1.69 [36] | 1.61 [14], 1.81[32], 0.85-1.63 [36] |
| This work | 1.60 | 1.50 | 1.70 |

The electronic structure of different phases has similarities. The valence band (VB) top states came from mainly of the p(I) orbitals, and the bottom of the conduction band (CB) predominantly from p(Pb) with a smaller contribution from the p(I) orbitals. It is in agreement with other references in the literature [14,37–39]. As a consequence the electronic properties for energies close to the bandgap are directly related to the $PbI_6$ octahedra. The contribution of the organic cation states near to bandgap is very small. They lie deep in the VB and CB. However, the size of the organic cation can distort the $PbI_6$ octahedra, and therefore, indirectly influence the electronic properties for energies near the bandgap. Furthermore, as we will show later, the contribution of the organic cation states to the absorption coefficient is appreciable for energies above the bandgap [19,20].

### 3.2 Absorption coefficient splitting

The results of the absorption coefficients for the C, T and O phases and its split in contributions (as a many-species expansion [19,20]) are shown in Figure 1 and Figure 2 respectively. They reproduce and allow the identification of the absorbance spectrum from reference [40]. The optical absorption band between 2.5-2.7 eV [40] (shaded in the figures) has been attributed to the transition from s(Pb) to p(Pb) [41]. It is in accordance with the results in the figures. This experimental band is due to the $\alpha_{Pb-Pb}$ intra-specie transition (mainly from of the s(Pb)-p(Pb)), and with a lower proportion of $\alpha_{I-I}$ (mainly from s(I)-p(I)), and $\alpha_{Pb-I}$ (p(Pb)-p(I)). Furthermore, the experimental absorption band between 3.2-3.5 eV [40] (shaded in the figures) has contributions, in descending order of importance, of the I-I, Pb-I and Pb-Pb species transition. The results in the figures also reproduce the decrease of the experimental absorption band between 3.5 and 4 eV [40].

Our results are also in accordance with the three peaks, at about 1.65, 2.20, and 3.10 eV, in the absorption spectra of the T perovskite structure in reference [42]. The first peak at about 1.65 eV is mainly due to the Pb-Pb intra-species excitations (predominantly s(Pb)-p(Pb)). The second peak, around 2.20 eV, is also mainly because of the Pb-Pb intra-species transitions, with a lower contribution of the I-I (s(I)-p(I)) excitation. The last peak, around 3.10 eV, is due to similar contributions from Pb-Pb and I-I intra-species, and Pb-I intra-species contributions (s(Pb)-p(Pb), s(I)-p(I), and p(Pb)-p(I) respectively).

The inter- and intra-species contribution of the organic cation atoms (C, H, N) is very small. Their contribution is mainly to the $A_{34}$ term. This term also has a high contribution similar to the Pb-Pb, I-I, and Pb-I contributions, but for higher energies. In fact, $A_{34}$ is one of the most important terms around 3.5 eV. The energy to which a transition between two states contributes to the absorption coefficient is the energy difference between the initial and final states of the transition. The electronic states of the organic cation atoms lie deep within the VB and CB whereas Pb and I have high contributions for energies close to the energy bandgap. Therefore the organic cation species contribution to the absorption coefficient (included in the $A_{34}$ term) is for larger energies than the bandgap, and larger than the Pb-Pb, I-I, and Pb-I contributions.

The contributions to the absorption coefficient (Figure 2) have a similar qualitative behavior for the three phases. However quantitatively the values are different. This difference is a consequence of the difference between the crystalline structures. It is clear that the electronic properties are primarily associated with the Pb-Pb, I-I, Pb-I, and $A_{34}$ contributions. As previously mentioned in the methodology section, changes in the bond lengths and angles in the different structures affect the electronic properties, and as a consequence the optical and the main photovoltaic parameters too.

### 3.2 Efficiency

As the generated photocurrent and the efficiency depend on the absorption coefficient (methodology section), different absorption coefficients will lead to different efficiencies. Therefore, the three perovskite structures analyzed, with different absorption coefficients (Figures 1 and 2), have different efficiencies. Using the previously obtained absorption coefficients the efficiencies have been calculated in accordance with the methodology section. From the results in Figure 3 the theoretical maximum efficiency for a $w=1$ μm-thick MAPI perovskite solar cell with a C-structure (with $Pm\overline{3}m$ symmetry) is $\eta \sim 26\%$, is in accordance with other results in the literature [43]. In this case, with $w \leq 1$ μm, the maximum efficiencies decrease from C→T→O symmetry as other results indicate in the literature [43]. However, for $w > 1.5$ μm the efficiency of the T-structure (the room temperature stable phase) is the largest, although the efficiency difference between the two phases is small, i.e. they did not change drastically across the T→C phase transition in accordance with the experimental results [9]. The O-structure is the phase with the lowest efficiency in all cases, although the efficiency difference with respect to the other phases is small. This also agrees with the experimental observation of a reduction in efficiency when approaching the orthorhombic phase [9].

Using a methodology similar to that used for the absorption coefficient splitting we have also split the efficiencies as an exact many-species expansion. Figure 4 shows the results of the most important inter- and intra-atomic contributions, as well as the total contribution of the terms of one and two species ( $A_{12}$ ), and those of 3 and 4 species ( $A_{34}$ ). The contribution percentage of the split efficiencies with respect to the total efficiency ( $(\eta_{split}/\eta_{Total}) \times 100$ ) is shown in Figure 5. The $A_{12}$ and $A_{34}$ contributions to the efficiency are similar for $w > 10$ μm. The most important $A_{12}$ contributions in descending order are from Pb-Pb, Pb-I and I-I. The rest of the $A_{12}$ contributions are almost zero, because they have an almost zero absorption coefficient in the AM1.5G spectra range of energies. The $A_{34}$ term also has a high contribution to the absorption coefficient (Figure 2) similar to the $A_{12}$ term, but for higher energies. Therefore, decreasing the width $w$ of the solar cell also decreases the $A_{34}$ contribution to efficiency. This is shown in the figure: for $w > 10$ μm the contributions to the efficiency of $A_{12}$ and $A_{34}$ are similar (around 50%) while for lower $w$ the contribution of $A_{34}$ decreases while that of $A_{12}$ increases.

## 4. Conclusions

The electronic, optical and photovoltaic properties of different MAPI crystalline structures are obtained from first principles based on the DFT. In all phases the VB and CB edges are predominantly made up of p(I) and p(Pb) states respectively. In order to identify, quantify and relate the structure with the optical and photovoltaic properties we have split these properties into a many-species expansion. This splitting is exact and there are no terms larger than those of 4 species.

The absorption coefficient of the different phases for energies close to the bandgap come mainly from the $A_{12}$ term contributions: Pb-Pb, and with a lower proportion of I-I, and Pb-I. The inter- and intra-species contributions of the organic cation atoms (C, H, N) to the $A_{12}$ term are very small. Their contribution is mainly to the $A_{34}$ term, similar in magnitude to the $A_{12}$ term but for higher energies. These contributions to the absorption coefficients are also reflected in the efficiencies of the different phases. Therefore the $A_{34}$ contribution to the efficiency decreases for thin solar cells whereas the $A_{12}$ increases, becoming the predominant contribution to the efficiency for ultra-thin solar cells.

The results justify the absence of abrupt changes observed experimentally in the optical and photovoltaic properties between different crystal structures.

**Acknowledgments**

This work has been supported by the National Spanish project MULTIPLER (RTI2018-096937-B-C21) and the European Commission through the funding of the project GRECO (H2020-787289).

**List of figures**

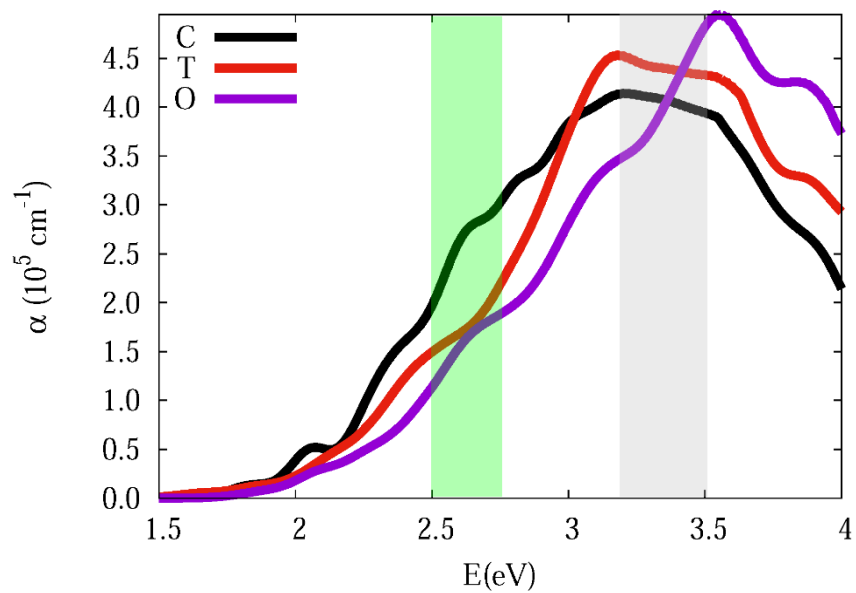

**Figure 1:** Absorption coefficient of the C, T, and O MAPI-structures as a function of the photon energy. The shaded areas correspond to the absorption peaks of the room temperature absorbance experimental spectrum from reference [40].

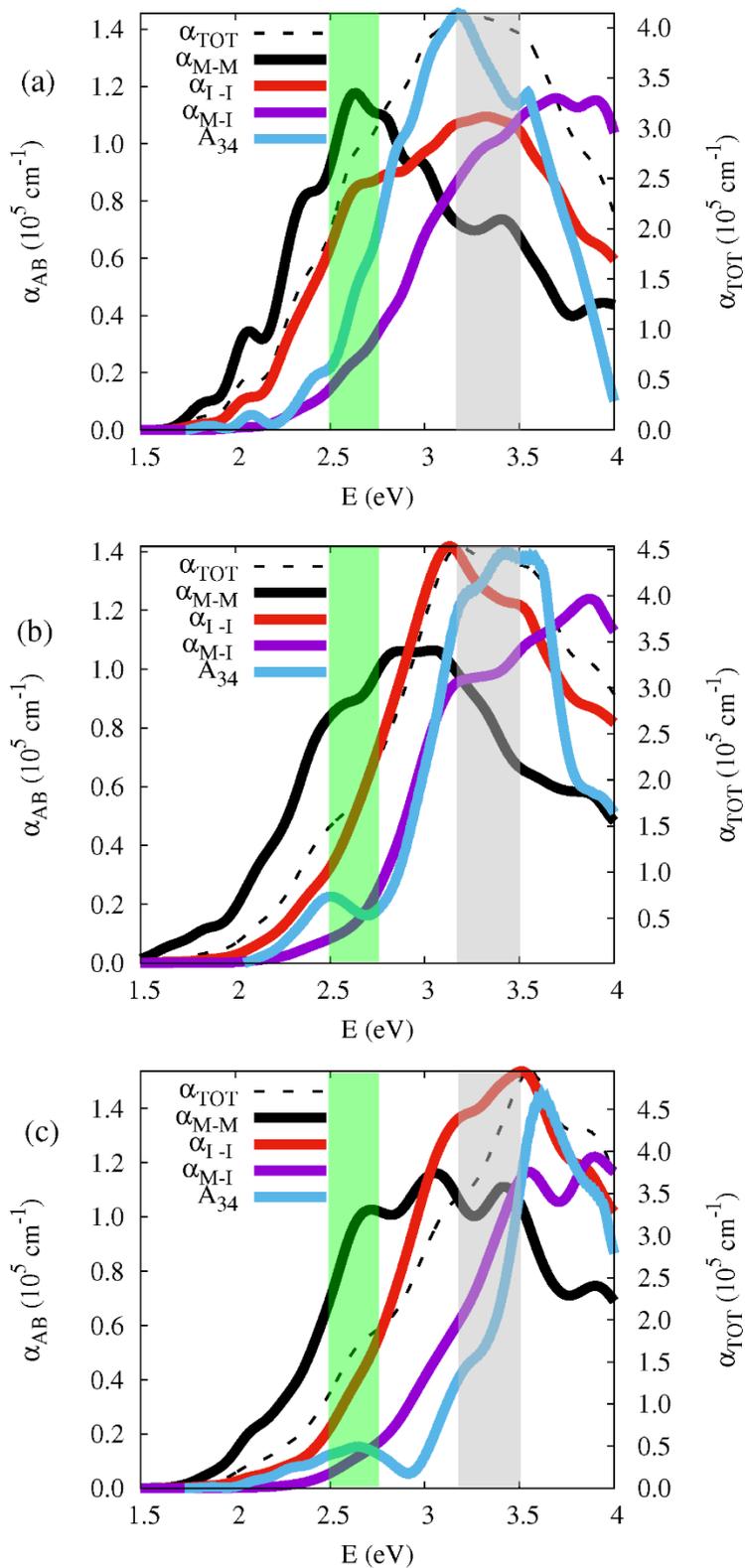

**Figure 2:** (left $y$ axis): absorption coefficient split into species contributions of the different MAPI structures: (a) C, (b) T, and (c) O. (Right $y$ axis): total absorption coefficient. The shaded areas correspond to the absorption peaks of the room temperature absorbance experimental spectrum from reference [40].

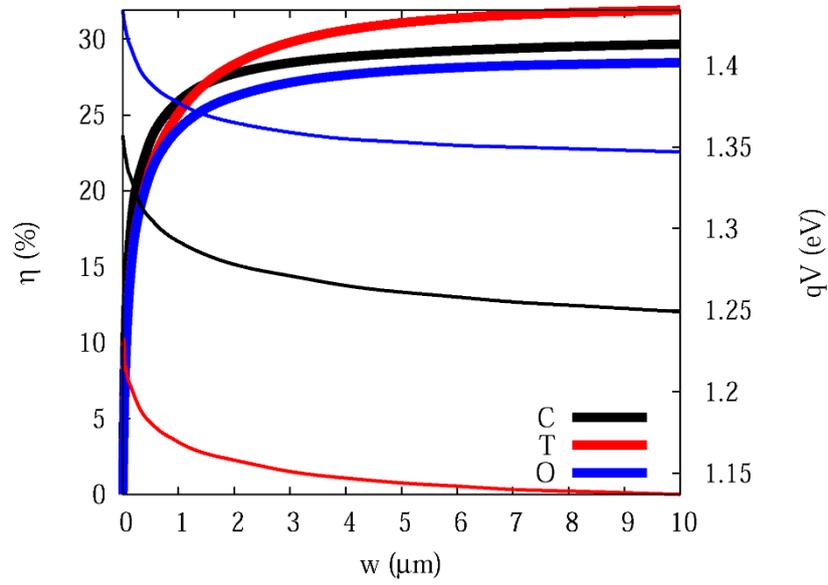

**Figure 3**: Efficiency $\eta$ (%) (left $y$ axis) and photo-voltage $V$ (right $y$ axis) for the maximum power point of the different MAPI structures as a function of the cell thickness $w$ using the AM1.5G spectra.

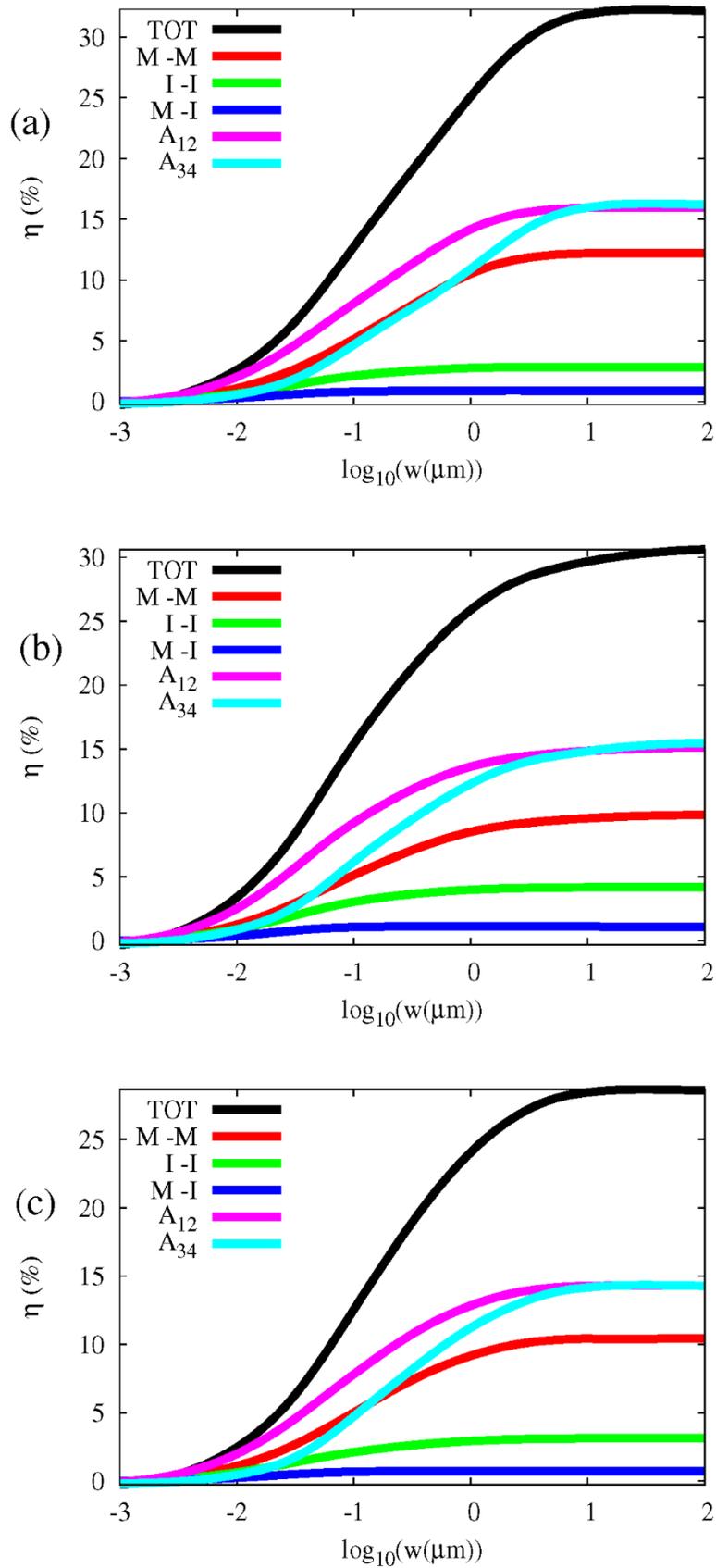

**Figure 4**: Efficiency $\eta$ (%) split into species contributions of the different MAPI structures: (a) C, (b) T, and (c) O as a function of the cell thickness $w$ decimal logarithm using the AM1.5G spectra.

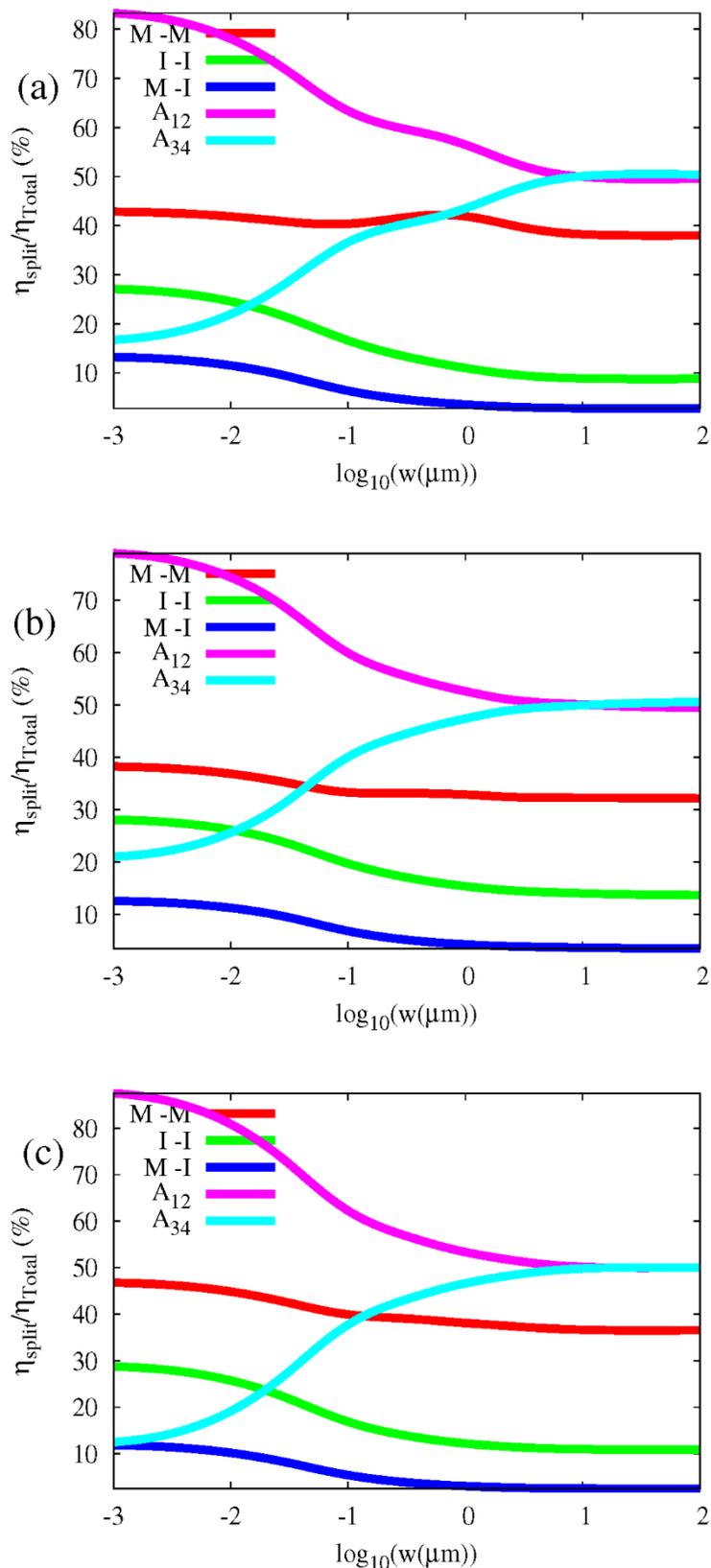

**Figure 5**: Percentage contribution of the split efficiencies with respect to the total efficiency $(\eta_{split}/\eta_{Total})\times 100$) split into species contributions of the different MAPI

structures: (a) C, (b) T, and (c) O as a function of the cell thickness $w$ decimal logarithm using the AM1.5G spectra.